\documentclass{elsart}
\input epsf.sty
\newcommand{\about}{$\sim \!$} 
\begin{document}  
\begin{frontmatter}
\title{ Development of a Laser Wire Beam Profile Monitor (I) }
\author{Yutaka Sakamura\thanksref{TIT}},
\author{Yasuo Hemmi\thanksref{DIT}},
\author{Hiroaki Matsuo\thanksref{DIED}},
\author{Hiroshi Sakai},  
\author{Noboru Sasao\thanksref{correspond}}
\address{Department of Physics, Kyoto University, Kyoto 606, Japan }
\author{Yasuo Higashi},
\author{Timo Korhonen\thanksref{PSI}},
\author{Takashi Taniguchi}, \\
 and 
\author{Junji Urakawa}
\address{High Energy Accelerator Research Organization (KEK), 
 Ibaraki 305, Japan }
\thanks[TIT]
 {Present address: {\it Physics Dept., 
         Tokyo Institute of Technology, Tokyo 152, Japan}}
\thanks[DIT]
 {Present address: {\it Daido Institute of Technology, Aichi 457, Japan}}
\thanks[DIED]
 {To the memory of HM who died on July 19, 1998.}
\thanks[correspond]
 {Corresponding author. e-mail: sasao@scphys.kyoto-u.ac.jp}
\thanks[PSI]
{Present address: {\it Paul Scherrer Institut, SLS,
               CH-5232 Villigen PSI, Switzerland}}

\journal{Nucl. Instr. and Meth. A}

\begin{abstract}

A conceptual design work and a basic experimental study 
  of a new beam profile monitor have been performed.
The monitor will be used 
 to measure emittance of an electron beam in the ATF damping ring at KEK,
 in which the transverse beam size of about $10 \,\mu {\rm m}$ is expected. 
It utilizes a CW laser and an optical cavity,
  instead of a material wire, to minimize
  interference with  an electron beam. 
A laser beam with a very thin waist is realized by 
  employing the cavity of nearly concentric mirror configuration
  while the intensity is amplified by 
  adjusting the cavity length to a Fabry-Perot resonance condition. 
We built a test cavity to establish a method to measure 
   important parameters such as a laser beam waist and 
   a power enhancement factor. 
Three independent methods were examined for the measurement of 
  the beam waist. 
It was found that the cavity realized the beam waist of 
   $20 \,\mu {\rm m}$ with the power enhancement factor of 50.

\vskip 1em
{\it PACS:\ }07.60.L, 41.75.H, 42.60
\end{abstract}
\end{frontmatter}

\section{Introduction}
Development of high energy $e^{+} e^{-}$ linear colliders
   is of crucial importance for the future particle physics.
Various R$\&$D works are now in progress to achieve 
  higher energy and higher luminosity.
Development of a high gradient RF cavity is essential to attain high energy
  while realization of a low emittance beam as well as a good  beam monitor 
  is important for high luminosity.
At KEK, the Accelerator Test Facility (ATF), consisting of 
  a 1.54 GeV linac and a damping ring, has been built 
  to study generation and manipulation of an ultra-low emittance 
  electron beam.
In accordance with this purpose, we have started developing 
  a beam profile monitor 
  to measure beam emittance in the ring.
Table \ref{table-DRbeam} shows the design parameters of the ring  
  relevant to our monitor~\cite{ref-ATF-design}.
Since the expected beam size is about  $10\,\mu $m
  vertically and about $ 60\,\mu$m horizontally,
  a beam profile monitor with 
  better than $ 10 \mu$m resolution is required.
A wire scanner made of tungsten or carbon is 
  one candidate~\cite{ref-wirescanner1}~\cite{ref-wirescanner2}.
However, a thin wire (for example, $10 \,\mu$m in diameter to 
  achieve a desired resolution) is expected to be destroyed due to 
  thermal stress caused by interactions with 
  the intense electron beam inside the ring.
Wire material will also influence the beam condition;
  an undesirable property for the monitor. 
  
Use of a laser beam, in stead of material wire, 
  has been proposed~\cite{ref-sintake}~\cite{ref-Ross}.
For example, a monitor using an intense pulsed laser was 
  tested successfully and achieved a resolution 
  in a sub-micron range~\cite{ref-sintake}. 
This monitor, however, works at a repetition
  rate of 10Hz and is not best suited to the quasi-continuous
  beam in the ATF damping ring; it would either take measurement time 
  very long or necessitate a powerful laser.

A new beam profile monitor is designed 
  using a CW laser and an optical cavity.
A laser beam with a very thin waist is realized by 
  employing the cavity of nearly concentric mirror configuration
  while the intensity is amplified by 
  adjusting the cavity length to a Fabry-Perot resonance condition. 
This monitor, referred to as a laser wire beam profile monitor, 
  operates as follows.
An electron interacts with laser light 
  and emits an energetic photon into the original 
  electron beam direction (Compton scattering).
The counting rate of a scattered photon is observed at various  
  wire positions; its shape then  
  gives a beam profile in one direction.
The monitor should be able to withstand an intense electron beam 
  without interfering it.
  
In this report, we describe a conceptual design of the beam profile monitor
   and some experimental results of a basic study 
   with a test cavity~\cite{ref-sakamura}.
The main aim of the study is to establish a  method to measure the size of 
  laser beam waist and the power enhancement factor. 
The paper is organized as follows : 
  Sec. 2 deals with a theoretical approach to the design of a 
  new monitor.
An experimental test setup and the results are shown in Sec.3. 
A summary and discussions are given in Sec.4.
     
\begin{table}
\begin{center}
\caption{Beam parameters of the ATF damping ring. 
 The values listed in the third column are used in the counting rate 
 calculation. Note($a$): These values indicate the beam sizes 
 at which this monitor will be installed.
        They are calculated from the designed emittance 
        of $1.4 \times {10}^{-9}$m$\cdot$rad (horizontal) and 
        $1.0 \times {10}^{-11}$m$\cdot$rad (vertical) with beta-functions of 
        ${\beta}_{x}=2.6$m (horizontal) and 
        ${\beta}_{y}=7.8$m (vertical), respectively.}
\vspace{3mm}
\begin{tabular}[b]{lccc} \hline \hline
 parameters                          & design value  & rate calculation & units  \\ \hline 
 beam energy                         & 1.54          & 1.54   & {\rm GeV}    \\ 
 circumference                       & 138.6         &        & {\rm m}  \\ \hline
 beam size (horizontal)${}^{(a)}$    & 61            & 61     & $\mu$m   \\ 
 beam size (vertical)${}^{(a)}$      & 8.8           & 8.8    & $\mu$m   \\ 
 beam size (longitudinal)            & 5             &        & {\rm mm} \\ \hline
 particles/bunch                     & 1 $\sim$ 3    & 1      & $10^{10}$   \\ 
 bunch spacing                       & 2.8 $\sim$ 5.6& 2.8    & {\rm nsec}  \\ 
 bunch /train                        & 10  $\sim$  60& 20     &           \\ 
 train spacing                       & 60            & 60     &{\rm nsec} \\ 
 train/ring                          &  2 $\sim$ 5   &  4     &         \\ \hline
 damping time (horizontal)           &  6.8          &        &{\rm msec}  \\ 
 damping time (vertical)             &  9.1          &        &{\rm msec}  \\
 damping time (longitudinal)         &  5.5          &        &{\rm msec}  \\ \hline \hline
\end{tabular}
\label{table-DRbeam}
\end{center}
\end{table}%

\section{Theoretical approach}

In this section, we consider a Compton process and an optical cavity.
We first present a relation between the energy and angle
  of the photon emitted by the Compton process.
We then calculate its cross section and an expected counting rate.
Several parameters, such as intensities and sizes of the electron and laser beam,
  must be assumed to do this calculation. 
We employ a set of parameters listed in the third column of Table~\ref{table-DRbeam}
  for the electron beam.
As to the laser beam, a 10 mW He-Ne laser and 
  an optical cavity with a beam waist of  $10 \,\mu$m
  and a power enhancement of 100 are supposed.
These cavity parameters are our goals at present.
We review a theory of an optical cavity in the latter half 
  of this section.

\begin{figure}[bhtp]
\hspace*{\fill}
\epsfysize=60mm 
\epsfbox{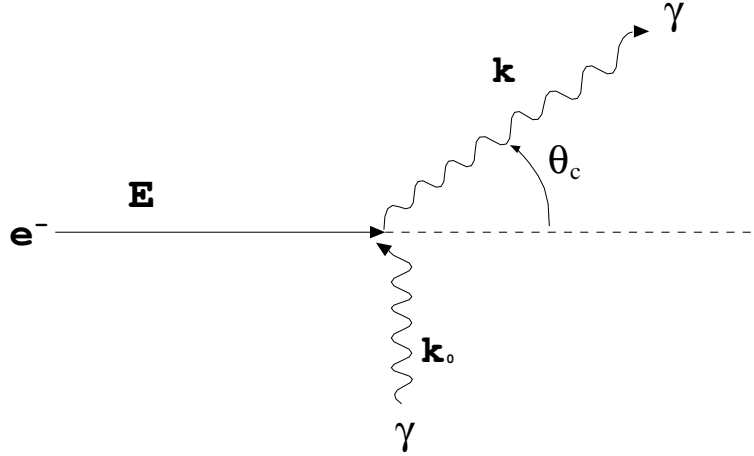} \hspace*{\fill}
\caption{Compton  scattering kinematics in the laboratory system. 
 (The recoil electron is not shown.)}
\label{fig-kinematics}
\end{figure}

\subsection{Compton scattering}
\begin{figure}[tbp]
\begin{center}
\epsfxsize=\textwidth
\epsfbox{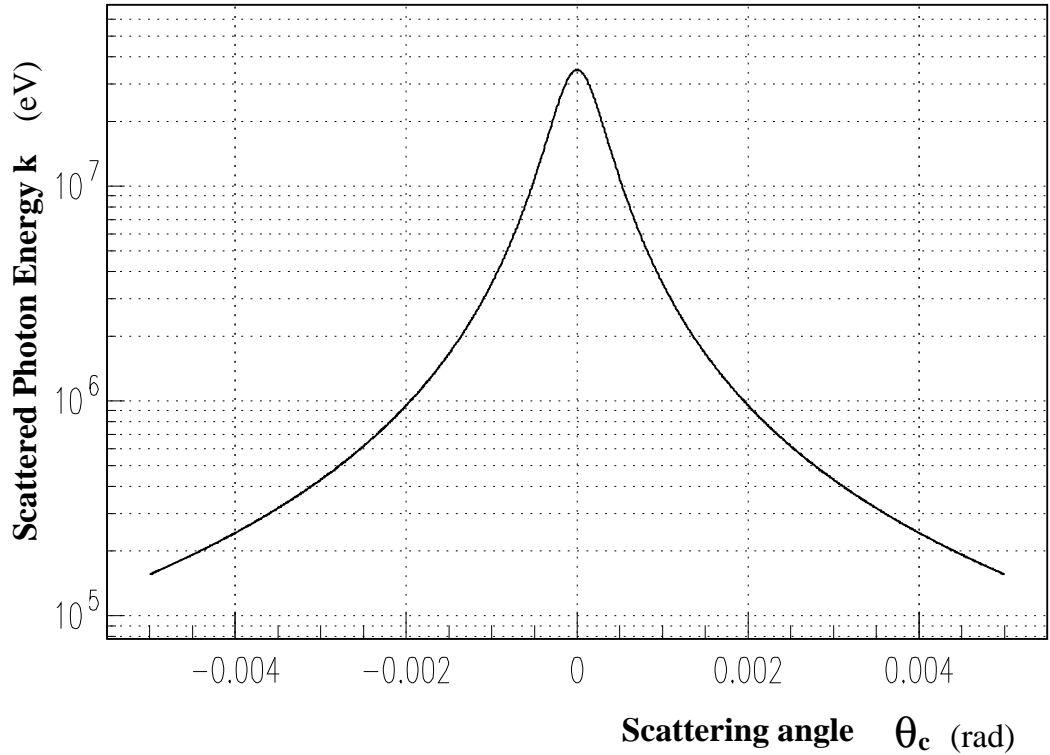}
\caption{ The energy $k$ of a scattered photon vs
          the scattering angle ${\theta}_{c} $.}
\label{fig-scattered}
\end{center}
\end{figure}

\paragraph{Kinematics}
Fig.\ref{fig-kinematics} shows the Compton scattering kinematics 
  in the laboratory system.
Here a laser light with an energy ${k}_{0}$ 
  is assumed to be injected perpendicular to 
  an electron beam with an  energy $E$.
The energy of scattered photon $k$ is given by
\begin{eqnarray*}
  k=\frac{k_{0} E}
    {E+k_{0}- \sqrt{E^2- m_{e}^2} \cos {\theta}_{c} }
\end{eqnarray*}
  where $m_{e}$ denotes the electron mass. 
In principle, $k$ depends upon  
  both polar angle ${\theta}_{c}$  and azimuthal angle
  ${\phi}_{c}$ (not shown in Fig.\ref{fig-kinematics}).
In practice, however, 
  $k$ depends only on ${\theta}_{c}$
  (hereafter referred to as a scattering angle),
   because the incident electron energy $E$ is
  much larger than the laser photon energy $k_{0}\,(E \gg k_{0})$.
The energy $k$ is plotted in Fig.\ref{fig-scattered} 
 as a function of ${\theta}_{c}$ for a He-Ne laser ($\lambda=633$nm).
We note that energetic  photons are emitted in the direction of 
 the incident electron;
 for example, photons with 10 MeV or larger are emitted within 
  $0.53$ mrad.
 
%

\begin{figure}[bhtp]
\begin{center}
\epsfxsize=0.95\textwidth 
\epsfbox{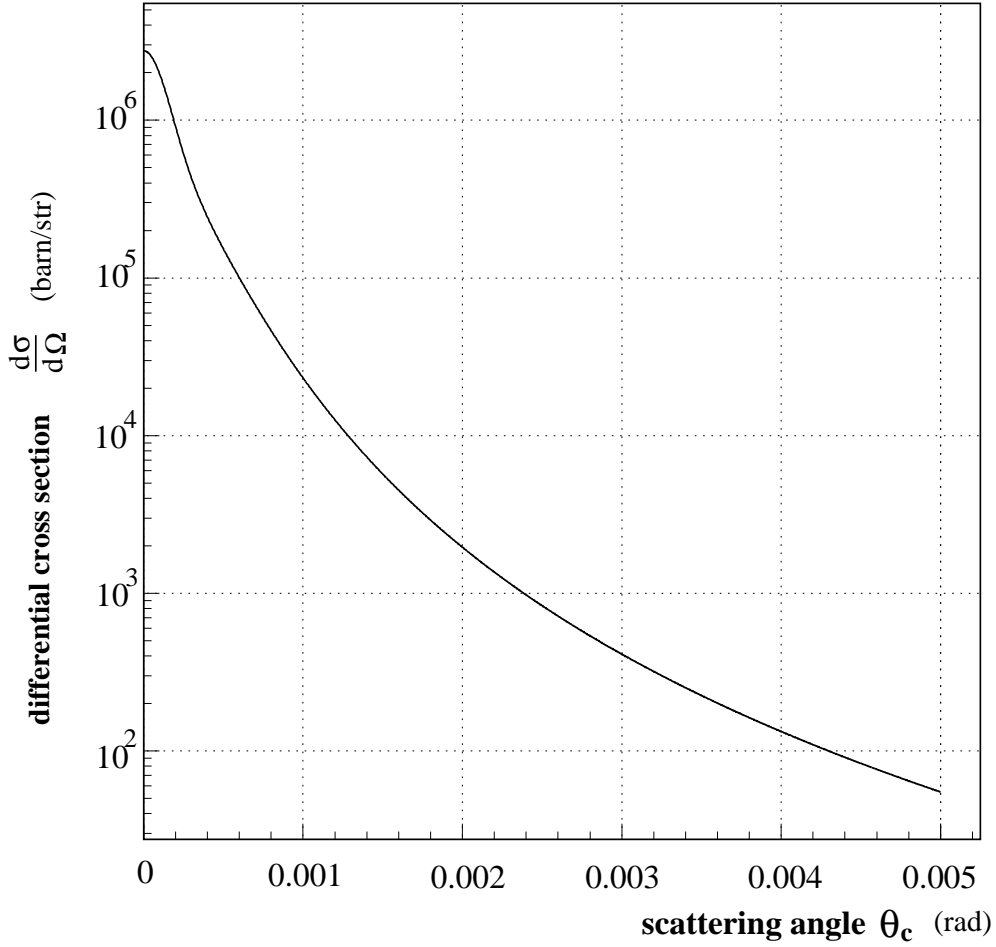}
\caption{ The Compton scattering cross section.}
\label{fig-xsection}
\end{center}
\end{figure}

\paragraph{Cross Section}
The cross section of the Compton process is given by 
 Klein-Nishina formula when the initial electron is at rest.
We made the appropriate Lorentz boost to calculate 
 the cross section in the laboratory frame.
The result is shown in Fig.\ref{fig-xsection}.
We note that the cross section is sharply peaked at ${\theta}_{c}=0$ :
  for example, the partial cross section 
  with ${\theta}_{c}< 0.53$ mrad ($k >10$ MeV) amounts to 
  0.44 barn, which should be compared with 
  the total cross section of 0.65 barn.
Evidently, to identify the Compton scattering unambiguously,
  it is best to detect the energetic photons emitted in the forward direction.
For the sake of argument, we assume that scattered photons with 
  $k >10$ MeV can be detected with 100\% efficiency in the following.

\paragraph{Counting rate}
Having determined the cross section of the Compton scattering, 
 we now estimate the counting rate. 
Here, as stated before,  we assume a 10 mW laser and an optical cavity with 
 a beam waist of $10\, \mu$m and a power enhancement factor of 100.
The counting rate is linear to the laser power and/or the enhancement
 factor so that extrapolation for other values is straightforward.
Taking all factors into account,
 we found the counting rate to be 3.2 kHz (horizontal measurement) 
and 28.8 kHz (vertical  measurement)~\cite{note-dia10}.
Here the horizontal (vertical) measurement represents the case in which 
   the electron's horizontal (vertical) beam size is measured
   at the peak of a gaussian-like beam 
   with a vertical (horizontal) laser wire. 
Since the vertical beam size is much smaller than
  that of horizontal one, the counting rate is larger accordingly.

\subsection{Optical cavity}

\begin{figure}[tbp]
\begin{center}
\epsfxsize=0.95\textwidth 
\epsfbox{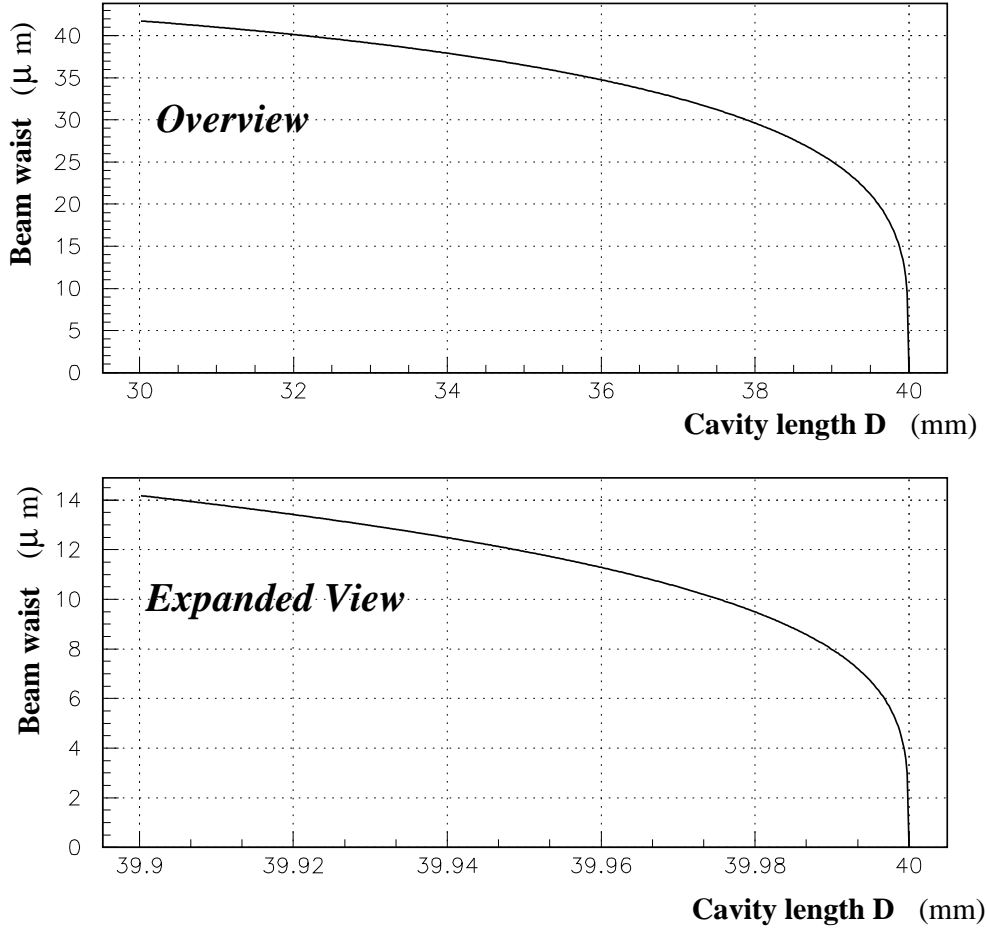}
\caption{ The beam waist $w_{0}$ as a function of the cavity length $D$.}
\label{fig-waist}
\end{center}
\end{figure}

In this section, we briefly summarize the theory of 
 an optical cavity (resonator) which is necessary for discussions 
 on the design and measurements of our test cavity.
It can be derived from the Helmholtz equation that 
  there exists a set of electromagnetic waves
  that have spherical wave fronts and Gaussian amplitudes.
These waves are called Gaussian beams~\cite{ref-gauss}.
The electric field of the fundamental TEM$_{00}$ mode is represented by
\begin{eqnarray*}
E_{0}(x,y,z)=A \frac{w_{0}}{w(z)} 
            \exp \left( -\frac{x^2+y^2}{w^2(z)} \right)
            \exp \left( i \frac{2 \pi}{\lambda} \cdot 
                                  \frac{x^2+y^2}{2R(z)} \right)
            \exp \left( i \Phi(z) \right),
\end{eqnarray*}
with
\begin{eqnarray}
   w(z)&=&w_{0} \sqrt{ 1 + \left( \frac{z}{z_{0}} \right)^2 }       
     \label{eq-beam-wz},                                            \\
   R(z)&=&z+\frac{z_{0}^2}{z},  \nonumber                             \\
   \Phi(z)&=& \arctan \left( \frac{z}{z_{0} } \right), \nonumber      \\
   z_{0}&=&\frac{\pi w_{0}^2 }{\lambda }               \nonumber     
\end{eqnarray}
 where $\lambda $ represents the wave length,
 $w(z)$ the beam spot size at the location $z$,
 $R(z)$ the curvature of the wave front,
 $\Phi(z)$ the Guoy phase factor,
 and $z_0$ the Rayleigh length.
The parameter $w_{0}$ is called the beam waist,
 and represents the smallest spot size realized at $z=0$.
The beam is well described by the geometrical optics where $|z|>z_{0}$.
Suppose two spherical mirrors with curvatures $R_{1}$ and $R_{2}$ are
 placed at $z_{1}$ and $z_{2}$, respectively.
If conditions
\begin{eqnarray*}
 R_{1}&=&R(z_{1})=z_{1}+z_{0}^2/z_{1} \\
-R_{2}&=&R(z_{2})=z_{2}+z_{0}^2/z_{2} 
\end{eqnarray*} 
 are satisfied, then the Gaussian beam can be stably confined in the optical
 cavity formed by the two mirrors.
It can be shown that there always exists a certain stable Gaussian beam
 for any curvatures $R_{1}$ and $R_{2}$
 and any cavity length $D$ ($D=z_{2}-z_{1}$), 
 if the stability condition
\begin{eqnarray*}
 0 \le \left( 1-\frac{D}{R_{1}} \right) \left( 1-\frac{D}{R_{2}} \right) \le 1
\end{eqnarray*}
 is satisfied. 
In other words, the properties of the stable Gaussian beam are uniquely determined
 once $R_{1}$, $R_{2}$ and $D$ are given.
In particular, the beam waist is represented by
\begin{eqnarray}
w_{0}^2=\frac{ \lambda }{ \pi }
        \frac{  \sqrt { D (R_{1}-D) (R_{2}-D) ( R_{1}+R_{2}-D )}  }
             { |    R_{1}+R_{2}-2 D   |                           }
        =\frac{ \lambda }{ \pi }
         \frac{ \sqrt { D (2 R-D )}  }{ 2 }
\label{eq-waist}
\end{eqnarray}
 where the latter equality holds for the mirrors with equal
 curvatures, i.e. $R=R_{1}=R_{2}$.
Fig.\ref{fig-waist} shows the beam waist $w_{0}$ as a
 function of $D$ for $\lambda=633$ nm and $R=20$ mm.
(These values of $\lambda$ and $R$ are employed in the actual studies  
  described in the next section.) 
It can be seen from the figure that, in order to realize a very thin
beam waist, the cavity must be nearly concentric ($D \simeq 2R$).
For example, a beam waist of 10 $\mu$m
 is realized when $D=40{\rm mm}-24 \mu{\rm m}$.
It should also be noted that the requirement for the setting 
accuracy in $D$ is very severe.

%

A stable Gaussian beam can be produced by injecting 
 an appropriate laser beam into a cavity.
The spot and divergence of the input laser beam at the mirror
 must match with those of the Gaussian beam in the cavity.
Some of the laser light is transmitted to the other side of
 the cavity.
The transmission ratio T is given by the Airy function represented by
\begin{eqnarray}
T=1/\left[
           1+\frac{4 {\it F}^2 }{ \pi^2 } 
            \cos^2  \left( \frac{2 \pi D}{ \lambda } -\delta \right)
          \right]
\label{eq-finesse}
\end{eqnarray}
 where $\delta $ is a phase shift which characterize the reflection at
 the mirrors.
The quantity $F$ is called ``finesse'' and is given by
\begin{eqnarray*}
F=\frac{ \pi \sqrt{R_{m}} }{ 1- R_{m} }
\end{eqnarray*}
where $R_{m}$ is the reflectivity of the mirrors.
  (The two mirrors are assumed to be identical here.)
Alternatively, the finesse $F$ may be expressed,
  from Eq.(\ref{eq-finesse}), by
\begin{eqnarray*}
F=\frac{ \Delta D (peak-to-peak) }{ \delta D (fwhm)  } 
 =\frac{ \lambda /2              }{ \delta D (fwhm)  }
\end{eqnarray*}
 where $\Delta D$ is the difference in $D$  between the two adjacent peaks,
 and $\delta D$ the full width at half maximum of a single peak.
The peak-to-peak distance $\Delta D$ is called the free spectral
 range, and is equal to a half of the wave length $\lambda$.

The power enhancement factor $P$ inside the cavity is given by
\begin{eqnarray*}
P=\frac{ T }{ 1- R_{m} }
  \left\{ 1+R_{m} -2 \sqrt{ R_{m} } \sin 
  \left( \frac{ \pi (z-D) }{ \lambda } + \delta \right)
  \right\}.
\end{eqnarray*}
It exhibits an interference pattern along the laser beam axis.
Averaging the above formula over $z$, 
  we obtain the average power enhancement factor $\bar{P}$:
\begin{eqnarray*}
\bar{P}=\frac{ 1+ R_{m} }{ 1- R_{m} }T \simeq \frac{2}{3} F T,
\end{eqnarray*}
which takes its maximum value of \, 
$\displaystyle
\frac{ 1+ R_{m} }{ 1- R_{m} }\simeq \frac{2}{3} F 
$ \,
when \, $T=1$.
For example, we need a mirror with reflectivity $R_{m}=98\%$ 
 in order to obtain a power enhancement factor of 100.


\section{Experimental Studies with a Test Cavity}
In this section, we describe our experimental studies 
with a test cavity. 
Their main purpose is to establish methods of
 measuring the relevant Gaussian beam parameters.
Specifically, we would like to measure the beam waist $w_{0}$ and
 the average power enhancement factor $\bar{P}$.
Since the beam waist is the most important parameter to our application,
we examined several independent methods of  measuring $w_{0}$.

%
\begin{figure}[tbp]
\epsfxsize=0.9\textwidth 
\epsfbox{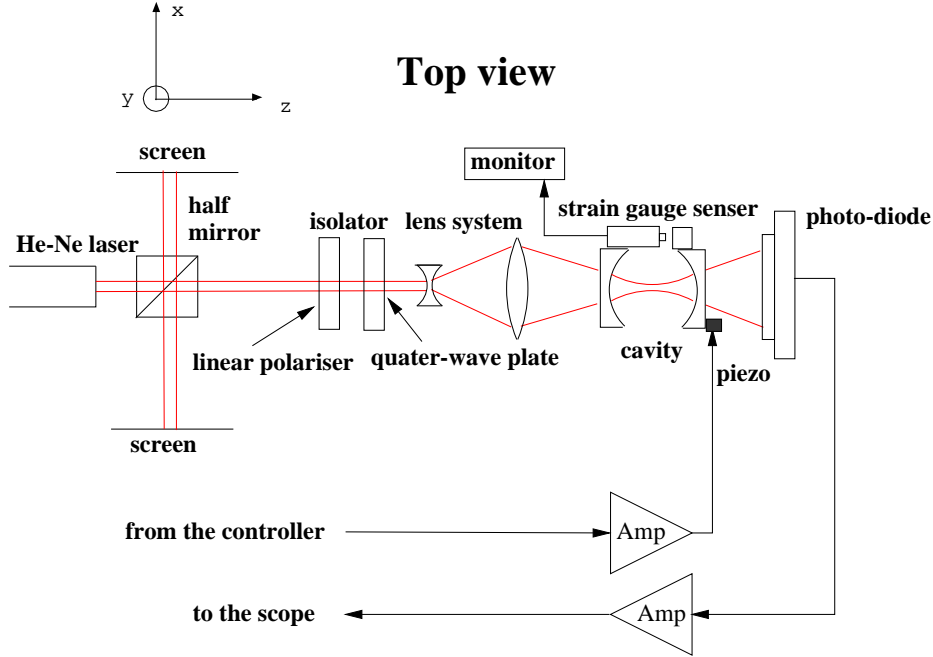}
\caption{ Schematic diagram of the test setup.}
\label{fig-setup}
\end{figure}

\subsection{Setup of the experiment}
Fig.\ref{fig-setup} shows a schematic diagram of our test setup.
It consists of a He-Ne laser ($\lambda=633$ nm)~\cite{ref-laser},
  an isolator, an input lens system,
  a test cavity, and a photodiode detector.
The laser provided a 1 mW beam with a diameter of 0.5 mm ($1/e^{2}$) and
 a full angular divergence of 1.6 mrad. 
The transverse spatial mode is ${\rm TEM_{00}}$.
In order to avoid a reflected beam to reenter the laser cavity, 
 an optical isolator consisting of a polarizer and a quarter-wave plate
 was placed at the laser exit.
A set of concave and convex lens formed an input lens system,
 which adjusted the laser beam to match with the Gaussian mode
 characteristics to the cavity.
We used two sets of spherical mirrors~\cite{ref-mirror} 
  of equal curvature ($R=20$mm) with different reflectivity.
One set had a nominal reflectivity of $96\%$ and the other set $85\%$.
The substrate was  BK7 glass ($n=1.519$): its spherical surface was coated
 with multi-layer dielectric materials and the other flat surface was polished.
Thus they acted as a concave lens for input and output light.  

These optical elements were installed on adjustable positioners.
In particular, the downstream mirror was staged on a piezo
 translator~\cite{ref-piezo} to scan the mirror along the beam direction $z$.
The piezo translator was controlled by a controller~\cite{ref-piezo} 
 and was monitored by a strain gauge sensor  
 with a 10 nm position resolution. 
We could set the piezo translator manually or scan it
 by an external voltage signal via the controller.
A PIN photodiode~\cite{ref-photodiode} was placed at the exit of the cavity.
Its output signal was fed into 
 a simple current-to-voltage amplifier 
 whose output was in turn monitored by an oscilloscope.
Unless otherwise noted, we tuned the cavity to the fundamental
  TEM$_{00}$ mode.

\subsection{Measurement of a power enhancement factor}

\begin{figure}[tbp]
\epsfxsize=0.9\textwidth 
\epsfbox{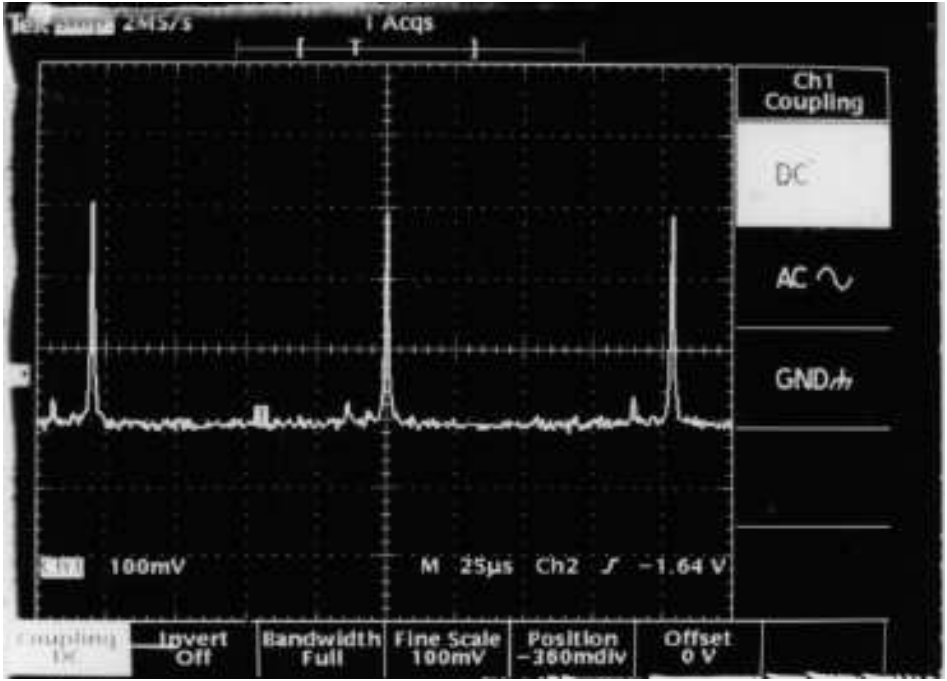}
\caption{ A typical example of the detector output, showing the Airy function.}
\label{fig-fsr}
\end{figure}

We  measured the finesse to evaluate a power enhancement factor.
A function generator was used to produce a triangle signal at
 a frequency of $\sim$100 Hz.
It was fed into the piezo controller.
Its amplitude was set so that the cavity length $D$ spanned over
 several free spectral ranges.
A typical example of the detector output is shown in Fig.\ref{fig-fsr}. 
We expected the horizontal trace to be  
  proportional to the change in the cavity length $D$
  since the piezo translator was driven by a triangle signal.
In reality, however, the relation was found to be non-linear due to
 a hysteresis effect of the piezo translator
 as well as a mechanical resonance effect of the mirror holder.
We measured on the oscilloscope the full width of a single peak 
 $(\delta D (fwhm))$ and the distances to 
 the flanking peaks  ($\Delta D $).
Then average was taken for the latter to cancel out the non-linear effect.  
%
%
%
Assuming that the observed $\Delta D$
 was equal to $\lambda  /2$, 
 we found the finesse $F$ to be 70 (22) 
 for the mirrors with $R_{m}=96\%  (85\%)$. 
The uncertainty in the measurement, stemming mainly from the 
 non-linear effect in $D$, 
 was estimated to be less than 10\%. 
The measured value should be compared with an expected value of $F$=77 (19).
We note that, for our actual application,
 precise knowledge of an absolute intensity is  unnecessary
 as long as it remains constant since a beam profile 
 would be measured as a relative shape of
 scattered photon's counting rate.

\subsection{Waist measurement by a shift-rotation method}

In order to obtain a beam waist of \about$10 \mu$m, 
 the cavity length should be set very close to twice
 of the mirror curvatures ($D \simeq 2R$). 
The tolerance in $D$ is rather severe; for example,
 to realize 10\% accuracy in the beam waist 
 ($\delta w_{0}/w_{0} = 10\%$) for $w_{0}=10 \mu$m, the setting error in $D$
 must be as small as \about$10 \mu$m.
This would be very difficult, if not impossible, to achieve
 since construction of the cavity necessarily induces setting errors,
 especially of glass mirrors.
Thus it is highly desirable to measure the cavity length $D$
 with sufficient accuracy.

\paragraph{Principle of shift-rotation method}
At first, we employed the following method~\cite{ref-sakamura}
 (referred to as a shift-rotation method).
Suppose that we have a cavity with unknown cavity length.
Suppose also that the cavity is tuned to the fundamental TEM$_{00}$ mode. 
Now we shift one of the mirrors laterally.
Since the mirrors are both spherical, one can always realign
   it by rotating the whole cavity.
A simple calculation shows that the displacement $ x_{\theta}$ 
and rotation angle
 $ \theta $
 are related by
\begin{eqnarray*}
 \tan (  \theta ) = \frac{ x_{\theta}}{2R-D}.
\end{eqnarray*}
Because the actual mirror is made of concave glass,
 it acts as a lens.
When this effect is taken into account,
the formula above should be replaced by
\begin{eqnarray}
 \sin (  \theta ) = n \sin 
 \left\{ \arctan \left( \frac{ x_{\theta}}{2R-D} \right) \right\}
\label{eq-shift-rotation}
\end{eqnarray}
where $n$ represents the refractive index of the glass substrate.
By measuring $ x_{\theta}$ and $ \theta$, 
we can determine $2R-D$, 
 and can calculate the beam waist $w_{0}$ with Eq.(\ref{eq-waist}).

%
\begin{figure}[tbp]
\hspace*{\fill} 
\epsfxsize=0.95\textwidth
\epsfbox{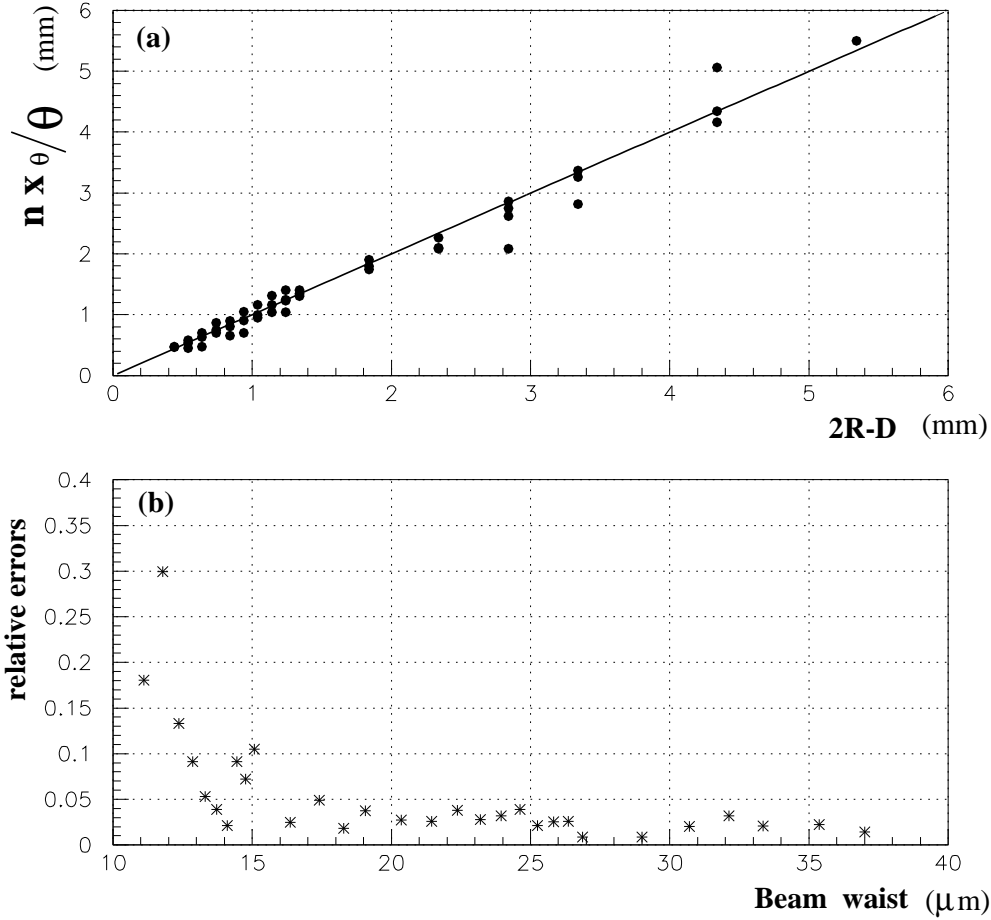} \hspace*{\fill}
\caption{ Beam waist measurement by shift-rotation method 
(a) $n {x}_{\theta}/\theta$ vs 2R-D. The solid line is a fit to 
 the data points by a straight line with a unit slope.
(b) $\delta {w}_{0}/{w}_{0}$ vs ${w}_{0}$. The beam waist ${w}_{0}$ and
its relative error $\delta {w}_{0}/{w}_{0}$ are obtained from measured 
$n {x}_{\theta}/\theta$ data and its rms deviation.}

\label{fig-Dcavity}
\end{figure}

\paragraph{Results of the measurement}
We measured several sets of $ x_{\theta}$ 
and $ \theta$ at a fixed cavity length $D$.
We then changed cavity length $D$ by moving the stage on which the
 downstream mirror was mounted.
We could trace the amount of a change in $D$ 
 because the stage was driven by a micrometer
 (attached to the stage in tandem with the piezo translator).
Fig.\ref{fig-Dcavity} (a) shows the results
  of such measurements.
The ordinate is the quantity
\begin{eqnarray}
 \frac{ x_{\theta}}{\tan \left\{ \arcsin \left( 
         { \sin  \theta }/{n}      \right) \right\}  }
 \simeq \frac{ n  x_{\theta} }{  \theta }
\label{eq-nx-over-theta}
\end{eqnarray} 
where the approximation holds for small $ \theta $.
The abscissa, labeled as $2R-D$, represents actually 
  $z$ position of the downstream mirror.
For a fixed z position,
each data point represents the measurement
 for different sets of $ x_{\theta}$ and $ \theta$.
It can be seen from Eq.(\ref{eq-shift-rotation}) that
 the quantity of Eq.(\ref{eq-nx-over-theta}) becomes $2R-D$.
Thus, ideally all the data points should reside on a straight
 line with a unit slope.
The solid line in the figure represents such a fit.
The intersection of the line with the abscissa is then relabeled as origin.
Having established the relation between $2R-D$ and the downstream mirror $z$ position,
 we calculated the beam waist $w_{0}$ with Eq.(\ref{eq-waist}).
We also calculated the root-mean-squared (rms) deviation of 
 the data points from the fitted line at fixed $2R-D$.
Each rms deviation was then converted to a relative error in $w_{0}$
 and is shown in Fig.\ref{fig-Dcavity} (b).
These errors originated from both $ x_{\theta}$ and $ \theta$ 
measurements:  the former was due mainly to the reading 
error of the micrometer
  and the latter to the uncertainty in determining the 
  exact $ \theta $ at which the realignment of the 
  optical axis was restored.
In this  paper,
  we assign a common error of $\delta w_{0}/w_{0}=0.05$
  to the measurements for  $w_{0}>20 \mu$m.
We would like to postpone drawing any conclusion
 for  $w_{0}<20 \mu$m~\footnote{
      This judgment was made partially because
      the cavity was found to be relatively unstable for 
      $w_{0} < 20 \mu$m.}.

\subsection{Measurement of the far field beam divergence}
It can be seen from Eq.(\ref{eq-beam-wz}) 
 that the beam width $w(z)$ in a far
 field is given by  
 $\displaystyle w(z) \simeq \frac{w_{0}}{z_{0}}z =
                      \frac{\lambda }{\pi w_{0}}z$.
Thus a measurement of the output laser profiles
 gives information on the beam waist $w_{0}$.
In particular, the beam intensity, when measured along $x$, reduces to
 $\displaystyle \frac{1}{\sqrt{e}}$ at 
 $\displaystyle x_{1/\sqrt{e}}=\frac{\lambda z}{2 \pi w_{0}}$
 from the peak value.

%
\begin{figure}[tbp]
\hspace*{\fill} 
\epsfxsize=\textwidth 
\epsfbox{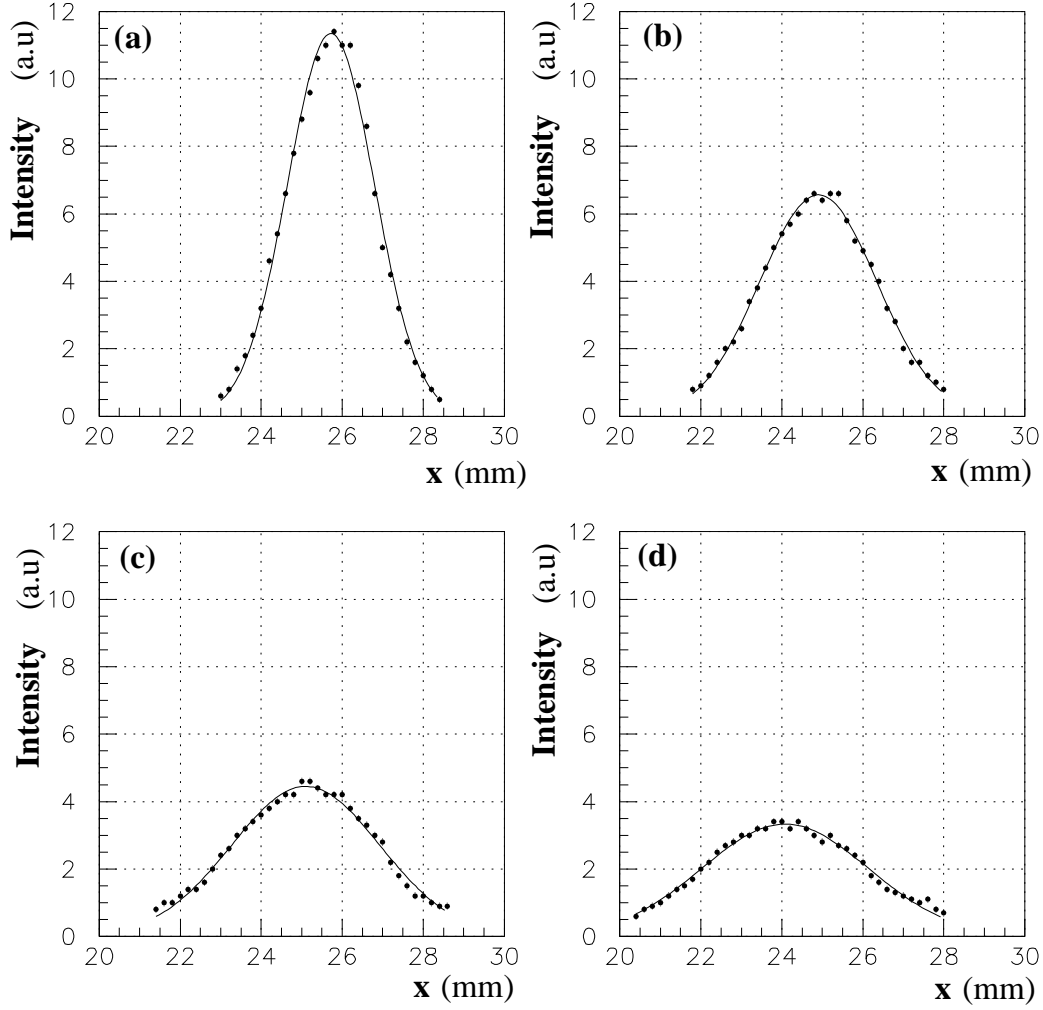} \hspace*{\fill}
\caption{ Intensity profile of the output laser beam at the positions 
of (a)\,$z={180}$, (b) 230, (c) 280, and (d) 330 mm 
 ($w_{0}=25 \mu$m). }
\label{fig-beamwidth25}
\end{figure}

\begin{figure}[tbp]
\hspace*{\fill} 
\epsfxsize=\textwidth
\epsfbox{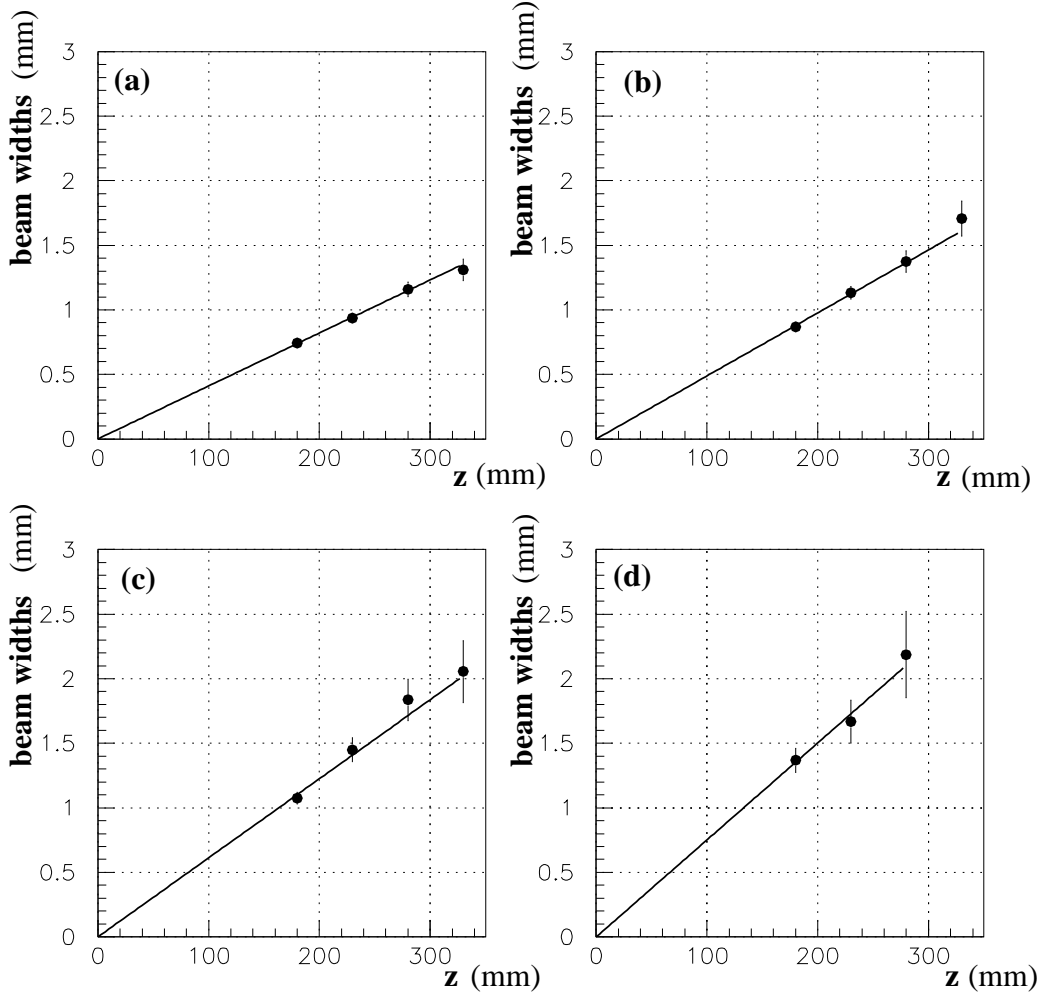} \hspace*{\fill}
\caption{ Measurements of the beam widths  
 $\displaystyle x_{1/\sqrt{e}} $ 
 as a function of $z$  for the nominal beam waist of 
 $w_{0}=$ (a) 35, (b) 30, (c) 25,  and (d) 20 $\mu $m. 
 The solid lines are the results of a straight line fit to the data points.}
\label{fig-beam-xvsz}
\end{figure}

Actual measurements were carried out for 4 different values of the
  beam waist; $w_{0}$=20, 25, 30 and 35 $\mu$m.
We call them ``nominal'' values because these were all determined by
 the shift-rotation method.
We inserted a slit with a horizontal width of $200 \mu$m
 in front of the photodiode.
It was then scanned horizontally ($x$ direction) 
  to measure the beam profile.
In order to enhance measurement accuracy, we determined the beam widths
 $\displaystyle x_{1/\sqrt{e}}$
 as a function of $z$, and obtained the beam waist $w_{0}$
 as a slope of these measurements~\footnote{
   In the actual analysis, the effect due to the concave mirror 
   (with refractive index $n$) is taken  into account.
   Then the beam width $\displaystyle x_{1/\sqrt{e}}$ 
   is given by 
   $\displaystyle x_{1/\sqrt{e}}=\frac{n \lambda z}{2 \pi w_{0}}$, 
   instead.}.
Fig.\ref{fig-beamwidth25} shows typical results of such a measurement 
 (for the nominal beam waist of $w_{0}=25 \mu $m).
Each solid line represents a Gaussian fit to the data points,
  from which the width $\displaystyle x_{1/\sqrt{e}}$ is determined.
Fig.\ref{fig-beam-xvsz} shows  $\displaystyle x_{1/\sqrt{e}} $
 as a function of $z$. 
Each straight line in the figures is  a linear fit to the data points and
 $w_{0}$ is  deduced from its slope.
%
%
%
%
Table \ref{table-beamwaist} shows the summary of the measurements
 (see the third column). 
The listed errors are all fitting errors.
Note that, since the nominal values are determined by the 
  shift-rotation method, they also have uncertainties,
  which are listed in the second column.
As seen, the two sets of the values agree well with each other.
(The fourth column will be explained in the next section.)

\subsection{Measurement of higher order  transverse modes}
The phase of a higher transverse mode ${\rm TEM_{mn}}$ is given by
\begin{eqnarray*}
 \phi (z) = -\frac{2 \pi}{\lambda} z + \left( m + n + 1 \right) \Phi(z)
\end{eqnarray*}
at $z$ on the beam axis.
These transverse modes interfere constructively when the phases 
 at the mirrors ($z = \pm D/2$) satisfy the resonance condition
\begin{eqnarray*}
 \phi (-D/2) - \phi (D/2) = p \pi 
\end{eqnarray*}
 where $p$ denotes an integer.
A straightforward calculation shows that
 the condition above is equivalent to 
\begin{eqnarray}
 D_{p}= \frac{\lambda }{2} 
       \left\{
              p+\frac{m+n+1}{\pi} \arccos 
              \left( \frac{D_{p}}{R}-1 \right) 
       \right\}.
\label{eq-higermode}
\end{eqnarray}
The transmitted beam intensity also exhibits its maximum value
 when the resonance condition is met.
Therefore, the spacing between the adjacent peaks 
 belonging to the same p is given by
\begin{eqnarray}
 \Delta D_{p}=D_{p}(m+n=1)-D_{p}(m+n=0)=\frac{\lambda}{2}
 \left[ \frac{1}{\pi} \arccos 
         \left( \frac{D}{R}-1 \right)  \right] 
\label{eq-highermode-spacing}
\end{eqnarray}
where the quantity $ D_{p}$ in the right hand side 
 of Eq.(\ref{eq-higermode}) is replaced by
 a representative value $D$ since its dependence is weak. 
Thus we can determine  $D$ by measuring the
 spacing $\Delta D_{p}$ of these peaks.
As represented in Eq.(\ref{eq-highermode-spacing}), we actually measured
 the spacing between the first excitation mode
 ($m+n=1$) and the fundamental mode ($m+n=0$).

\begin{table}[tbp]
\begin{center}
\caption{Summary of the beam waist measurements.
 The nominal values are defined to be those  
 determined by the shift-rotation method. Their uncertainties are 
 listed in the second column. }
\vspace{3mm}
\begin{tabular}{cccc} \hline
 nominal $w_{0}$  & shift-rotation 
           & $a_{1/\sqrt{e}}$ measurement &   transverse mode    \\ \hline
 20.0 & $\pm$ 1.0 $\mu$m  & 20.4 $\pm$ 1.7 $\mu$m & 20.1 $\pm$ 0.58 $\mu$m  \\ 
 25.0 & $\pm$ 1.3 $\mu$m  & 24.8 $\pm$ 1.2 $\mu$m & 24.9 $\pm$ 0.42 $\mu$m  \\ 
 30.0 & $\pm$ 1.5 $\mu$m  & 31.0 $\pm$ 1.6 $\mu$m & 30.0 $\pm$ 0.30 $\mu$m  \\ 
 35.0 & $\pm$ 1.8 $\mu$m  & 37.5 $\pm$ 1.4 $\mu$m & 36.4 $\pm$ 0.25 $\mu$m  \\ \hline
\end{tabular}
\label{table-beamwaist}
\end{center}
\end{table}

\begin{figure}[tbp]
\epsfxsize=0.9\textwidth 
\epsfbox{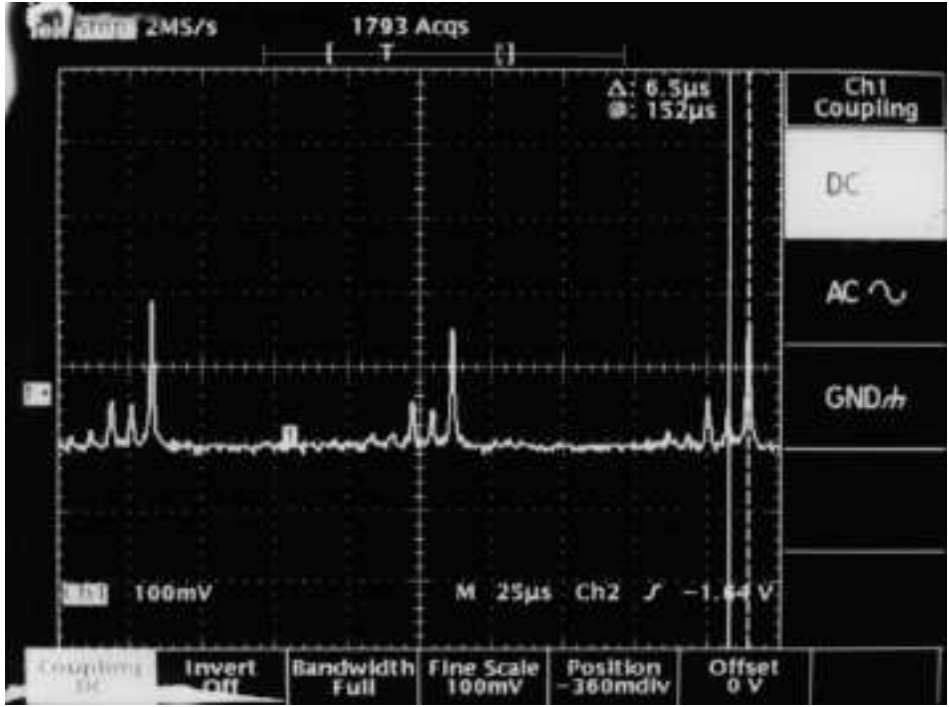}
\caption{An example of an output including higher transverse modes.}
\label{fig-higermode}
\end{figure}

After confirming the cavity to be in the fundamental
 TEM$_{00}$ mode, we detuned the 
 optical axis by shifting one of the mirrors laterally
 to excite the transverse mode.
Fig.\ref{fig-higermode} shows a typical example of 
 an output including higher transverse modes. 
We determined the spacing between the main peak (the fundamental mode) 
 and its adjacent peak (the first excitation mode) 
 as well as the spacing between two main peaks (a free spectral range).
Then we deduced the cavity length $D$ via Eq.(\ref{eq-highermode-spacing}).
The results of the measurements by this method are listed
 in the fourth column of Table \ref{table-beamwaist}.
Uncertainty came mainly from reading errors on the oscilloscope
 and also from non-linearity in $D$. 

%

\section{ Summary and Discussions}
We described in this paper a conceptual design
 of a beam profile monitor for an electron beam
 with a transverse size of \about $60 \mu$m (horizontal) 
and \about $10 \mu$m (vertical).  
Specifically, the monitor is intended to diagnose 
 an electron beam in the ATF damping ring at KEK.
The monitor works as follows: we realize a very thin
 (\about  $10 \mu$m) laser wire in an optical 
 cavity.
Energetic photons are emitted by the Compton scattering
in the direction of the electron beam.
The counting rate, measured as a function of 
 the laser beam position, will  give information on
 the electron beam profile.

We calculated the scattered photon energy spectrum 
 (see Fig.\ref{fig-scattered}) and the Compton cross
 section (see Fig.\ref{fig-xsection}) for the 1.54 GeV
 electron beam.
Assuming a 10 mW laser and an optical cavity with a beam waist of $10 \mu$m 
 and a power enhancement of 100, 
 the expected counting rate is 3.2 kHz (horizontal) 
and 28.8 kHz (vertical) when scattered 
 photons with energies $>$10 MeV are detected.

A key element of the proposed monitor is the optical cavity,
 in which a Gaussian beam with a $10 \mu$m beam waist must be 
 realized.
A power enhancement inside the cavity is an another 
 important parameter to investigate.
We thus constructed a test cavity to measure these parameters.
The cavity realized the power enhancement of 50 and the beam
 waist of $20\mu$m.
For the beam waist measurement, we employed three independent methods. 
As seen in Table \ref{table-beamwaist}, 
 the results obtained by these methods
 are consistent with each other.
The relative error $\delta w_{0}/w_{0}$ is about 5\% by
 the shift-rotation method, 
 less than 10\% by the beam divergence measurement 
 and less than  3\% by the higher transverse mode measurement.
We measured the finesse $F$ to obtain a power enhancement
 factor.
Using the mirrors with reflectivity of $ R_{m}=96\%$,
 we obtained the enhancement of 50 ($F=70$), 
 which is consistent with the expected value.

Several comments are in order here.
First,
 we have not yet demonstrated a $10 \,\mu {\rm m}$ beam waist.
The main defect in this test cavity is its mechanical rigidity.
In particular the mirror holder (supporting rod)
 caused disturbing resonant vibration.
This can, however, be overcome with a proper cavity design.
The second comment is on comparison between the three methods 
 employed to measure the beam waist.
The measurement of the beam width $x_{1/\sqrt{e}}$ is 
 very simple and reliable. 
A newly devised method, the shift-rotation method, gave  more
 accurate results than the beam divergence measurement. 
However, these methods would not be applicable for an actual
 cavity installed in an electron beam line.
The third one, the observation of higher transverse modes, 
 is simple and the most accurate among the three.
Thus this method is best suited to an in-situ measurement.
Finally, we comment on a power enhancement factor or 
  an effective power inside the cavity.
We demonstrated the enhancement factor of 50.
Obviously a higher power is desirable: it would allow us 
 faster measurement and would enable us  
 to study, for example, an electron cooling process and/or a 
 size of an individual bunch.
In this regard, we plan to employ a laser with a higher power and 
 mirrors with a higher reflectivity for a prototype monitor.
A design work is now underway.

\begin{ack}
It is our pleasure to thank Dr. M. Ross for suggesting this work 
and Prof. H. Sugawara and K. Takata
for their support and encouragement.
\end{ack}


\end{document}